\documentclass[12pt]{article}
\usepackage{graphics}
\usepackage{bm}
\usepackage{graphicx}
\usepackage{amsmath}
\usepackage{amssymb}
\usepackage{amstext}
\usepackage{amscd}
\usepackage{amsfonts}
\usepackage{appendix}
\usepackage{hyperref}
\usepackage{epstopdf}
\usepackage{color}
\DeclareMathAlphabet{\EuFrak}{U}{euf}{m}{n}
\DeclareMathAlphabet{\EuScript}{U}{eus}{m}{n}

\newcommand{\nd}{\noindent}

\newcommand{\be}{\begin{equation}}
\newcommand{\ee}{\end{equation}}
\newcommand{\ben}{\begin{eqnarray}}
\newcommand{\een}{\end{eqnarray}}

\title{{\bf Spatial cut-offs, Fermion Statistics, and  Verlinde's Conjecture}}
\author{{\small{A. Plastino$^{1,3,4}$, M. C. Rocca$^{1,2,3}$}}, \\
\small{$^1$ Departamento de F\'{\i}sica,
Universidad Nacional de La Plata,}\\
\small{$^2$ Departamento de Matem\'{a}tica,
Universidad Nacional de La Plata,}\\
\small{$^3$ Consejo Nacional de Investigaciones Cient\'{\i}ficas
y Tecnol\'{o}gicas}\\
\small{(IFLP-CCT-CONICET)-C. C. 727, 1900 La Plata -
Argentina}\\\small{$^4$  SThAR - EPFL, Lausanne, Switzerland}}

\date{\today}

\begin{document}

\maketitle

\begin{abstract}
\nd Verlinde conjectured eight years ago that gravitation might be an emergent entropic force.
This rather surprising assertion was proved in [Physica A {\bf 505} (2018) 190] within a purely
classical statistical context, and in [DOI: 10.13140/RG.2.2.34454.24640] for the case of bosons' statistics. In the present work, we appeal to a quantum scenario involving fermions' statistics. We  consider also the classical limit of quantum
(statistical)  mechanics (QM).   
\nd We encounter a lower bound to the distance $r$ between the two interacting
masses,     i.e., an $r$ cut-off. This is a new effect that exhibits some resemblance with the idea of space
discretization proposed by recent gravitation theories.\\
\vskip 3mm \nd {\bf Keywords}: Gravitation, fermions, entropic force, emergent force, Verlide's conjecture.

\end{abstract}

\newpage

\tableofcontents

\newpage

\renewcommand{\theequation}{\arabic{section}.\arabic{equation}}

\section{Introduction}

\setcounter{equation}{0}

Eight years ago, Verlinde \cite{verlinde} proposed to establish a bridge linking gravity with an entropic force. The ensuing conjecture was proved recently I) in a purely classical environment \cite{p1} 
and  II) in \cite{DOI} for the Bose-Einstein statistics.\vskip 3mm

\nd According to Verlinde, gravity should emerge as a result of information about the positions of material particles, connecting a thermal treatment of gravity to 't Hooft's holographic principle. In this view, gravitation could be regarded as  an emergent phenomenon. This interesting  Verlinde's notion was the focus of ample attention. See for
instance \cite{times,libro,prd}. A very good overview on the gravitation's statistical mechanics is that of   Padmanabhan \cite{india}, and references therein.\vskip 2mm

\nd Verlinde's idea motivated works on cosmology, the dark energy hypothesis, cosmological acceleration, cosmological inflation, and loop quantum gravity. The pertinent literature is abundant \cite{libro}.  An important contribution was made by  Guseo \cite{guseo}, who demonstrated that the local entropy function, related to a
logistic distribution, is a catenary and vice versa, an invariance that can 
be interpreted in terms of   Verlinde’s gravity's origin conjecture. 
  Guseo puts forward a new  interpretation of the local entropy in a system  \cite{guseo}.
Recapitulating: 
\nd  Verlinde's conjecture has been proved:
 \begin{itemize}
 \item In a classical scenario in \cite{p1}.\\
 \item In a quantum environment for bosons in ref.
 \cite{DOI}.
 \end{itemize} 
\nd In this paper we wish to address the Verlinde-associated  quantum fermionic case, 
that will  be seen to    yield rather surprising results.

\section{Entropic force for  an $N-$particles Fermi gas }
 
\nd We base our considerations on Chapter 6  of \cite{lemons}, where the reader is referred to  for details. It is  assumed
that each fermion possesses an  average energy $E/ N
$. Such average energy approximation produces results that, while
approximate, describe important features of the ideal Fermi gas \cite{lemons}.

\subsection{Quantum entropic force}

\setcounter{equation}{0}

Our $N-$fermions  gas' entropy can be written in the fashion  [see \cite{lemons}, Eq. (6.15)]
\begin{equation}
\label{eq2.1}
S=Nk_B\left[\ln\left(\frac {n} {N}-1\right)-
\left(\frac {n} {N}\right)\ln\left(1-\frac {N} {n}\right)\right],
\end{equation}
where $n$ is related to the system's energy $E$ according to \cite{lemons}
\begin{equation}
\label{eq2.2}
n=V\left(\frac {E} {N}\right)^{\frac {3} {2}}
\left(\frac {4\pi e m} {2h^2}\right)^{\frac {3} {2}}.
\end{equation}
We can cast the volume as that of a sphere 
\begin{equation}
\label{eq2.3}
V=\frac {4} {3}\pi r^3,
\end{equation}
whose area is to be called $A$. Then we can recast  (\ref{eq2.1})    as

\[S=Nk_B\ln\left[1-\frac {V} {N}\left(\frac {E} {N}\right)^{\frac {3} {2}}
\left(\frac {4\pi em} {2h^2}\right)^{\frac {3} {2}}\right]-\].
\begin{equation}
\label{eq2.4}
k_BV\left(\frac {E} {N}\right)^{\frac {3} {2}}
\left(\frac {4\pi em} {2h^2}\right)^{\frac {3} {2}}
\ln\left[\frac {N} {V}\left(\frac {N} {E}\right)^{\frac {3} {2}}
\left(\frac {2h^2} {4\pi em}\right)^{\frac {3} {2}}-1\right]
\end{equation}
Now, according to \cite{p1}, the entropic force $F_e$ is obtained via derivative with respect to $A$
\[F_e=-\lambda\frac {\partial S} {\partial A}=\]
\[\frac {\lambda 3k_BN} {8\pi r^2}
\frac {1} {\frac {3N} {4\pi r^3}\left(\frac {N} {E}\right)^{\frac {3} {2}}
\left(\frac {3h^2} {4\pi e m}\right)^{\frac {3} {2}}-1}+\]
\[\frac {\lambda k_B} {2}\left(\frac {E} {N}\right)^{\frac {3} {2}}
\left(\frac {4\pi e m} {2h^2}\right)^{\frac {3} {2}} r
\ln\left[\frac {3n} {4\pi r^3}\left(\frac {N} {E}\right)^{\frac {3} {2}}
\left(\frac {3h^2} {4\pi e m}\right)^{\frac {3} {2}}-1\right]-\]
\begin{equation}
\label{eq2.5}
\frac {\frac {\lambda k_B} {2}\left(\frac {E} {N}\right)^{\frac {3} {2}}
\left(\frac {4\pi e m} {2h^2}\right)^{\frac {3} {2}} r}
{\frac {4\pi r^3} {3N}\left(\frac {E} {N}\right)^{\frac {3} {2}}
\left(\frac {4\pi e m} {2h^2}\right)^{\frac {3} {2}}-1},
\end{equation}
where $\lambda$ is an arbitrary constant. We now recast the above equations  as
\[F_e=\frac {12\lambda k_BN\left(\pi emE\right)^{\frac {3} {2}} r}
{32\pi r^3\left(\pi emE\right)^{\frac {3} {2}}-
3^{\frac {5} {2}}N^{\frac {5} {2}}h^3}-\]
\[\frac {4\pi\lambda k_B\left(\pi emE\right)^{\frac {3} {2}}}
{\left(3N\right)^{\frac {3} {2}}h^3}r
\left\{\ln\left[32\pi r^3\left(\pi emE\right)^{\frac {3} {2}}-
(3N)^{\frac {5} {2}}h^3\right]-
\ln\left[32\pi r^3\left(\pi emE\right)^{\frac {3} {2}}\right]\right\}-\]
\begin{equation}
\label{eq2.6}
\frac {12\lambda k_BN\left(\pi emE\right)^{\frac {3} {2}} r}
{32\pi r^3\left(\pi emE\right)^{\frac {3} {2}}+
(3N)^{\frac {5} {2}}h^3}.
\end{equation}
Finally, we arrive at 
\begin{equation}
\label{eq2.7}
F_e=\frac {4\pi\lambda k_B\left(\pi emE\right)^{\frac {3} {2}}}
{\left(3N\right)^{\frac {3} {2}}h^3}r
\left\{\ln\left[32\pi r^3\left(\pi emE\right)^{\frac {3} {2}}-
(3N)^{\frac {5} {2}}h^3\right]-
\ln\left[32\pi r^3\left(\pi emE\right)^{\frac {3} {2}}\right]\right\},
\end{equation} our central result here.

\subsection{Fermionic entropic force in the classical limit (CL)}

\nd The CL obtains for \cite{lemons}
\begin{equation}
\label{eq2.8}
\frac {N} {n}<<1,
\end{equation}
and in this limit the entropy becomes \cite{lemons}
\begin{equation}
\label{eq2.9}
S=Nk_B\left[1+\ln\left(\frac {n} {N}\right)\right],
\end{equation}
or
\begin{equation}
\label{eq2.10}
S=
\frac {5Nk_B} {2}+Nk_B\ln\left[\frac {V} {N}\left(\frac {E} {N}\right)^{\frac {3} {2}}
\left(\frac {4\pi m} {2h^2}\right)^{\frac {3} {2}}\right].
\end{equation}
Now we have an entropic force of the form
\begin{equation}
\label{eq2.11}
F_e=-\lambda\frac {\partial S} {\partial A}=
-\frac {\lambda 3Nk_B} {8\pi r^2},
\end{equation}
which is indeed of the Newton appearance, so that Verlinde's conjecture gets proved in the classical limit.
 Note also that the entropic force (\ref{eq2.11}) can be derived as well from (\ref{eq2.7}) by taking $r$ large enough. Using now the equality
\begin{equation}
\label{eq2.12}
-\frac {\lambda 3Nk_B} {8\pi r^2}=
-\frac {GmM} {r^2},
\end{equation}
where  $G$ is the gravitational constant, we see that $\lambda=\lambda (m,M,N)$  in the case
$r$ large enough.  

\subsection{Entropic Potential Energy}

The entropic force is proportional to the derivative of the entropy with respect to 
the area $A$ of the sphere.  It is interesting  to calculate 
the corresponding potential energy $E_P$. For this we define the constants
$a=(3N)^{\frac {5} {2}}h^3$ and $b=32\pi(\pi e mE)^{\frac {3} {2}}$.
Using  reference \cite{gra} we can compute the potential energy from the expression of the 
entropic force. The ensuing calculation is simple but lengthy. Its result reads 
\[E_P(r)=-\frac {3Nk_B\lambda} {8\pi}\frac {b} {a}\left\{\frac {r^2} {2}
\ln\left(1-\frac {a} {br^3}\right)-
\frac {a^{\frac {2} {3}}} {2b^{\frac {2} {3}}}\left\{\frac {1} {2}\ln
\left[\frac {\left[r-\left(\frac {a} {b}\right)^{\frac {1} {3}}\right]^2} 
{r^2+\left(\frac {a} {b}\right)^{\frac {1} {3}}r+\left(\frac {a} {b}\right)^{\frac {2} {3}}}
\right]\right.\right.+\]
\begin{equation}
\label{eq2.13}
\left.\left.\sqrt{3}\left[\arctan\left[\frac{2r+\left(\frac {a} {b}\right)^{\frac {1} {3}}}
{\sqrt{3}\left(\frac {a} {b}\right)^{\frac {1} {3}}}\right]
-\frac {\pi} {2}\right]
\right\}\right\}
\end{equation}
where we have selected $E_P(r)=0$ when $r\rightarrow\infty$
For $r$ large the potential energy adopts the appearance
\begin{equation}
\label{eq2.14}
E_P(r)=
-\frac {\lambda 3Nk_B} {8\pi r},
\end{equation}
which is coherent with the result  (\ref{eq2.11}).

\subsection{Results}

\begin{figure}[h]
\begin{center}
\includegraphics[scale=0.5,angle=0]{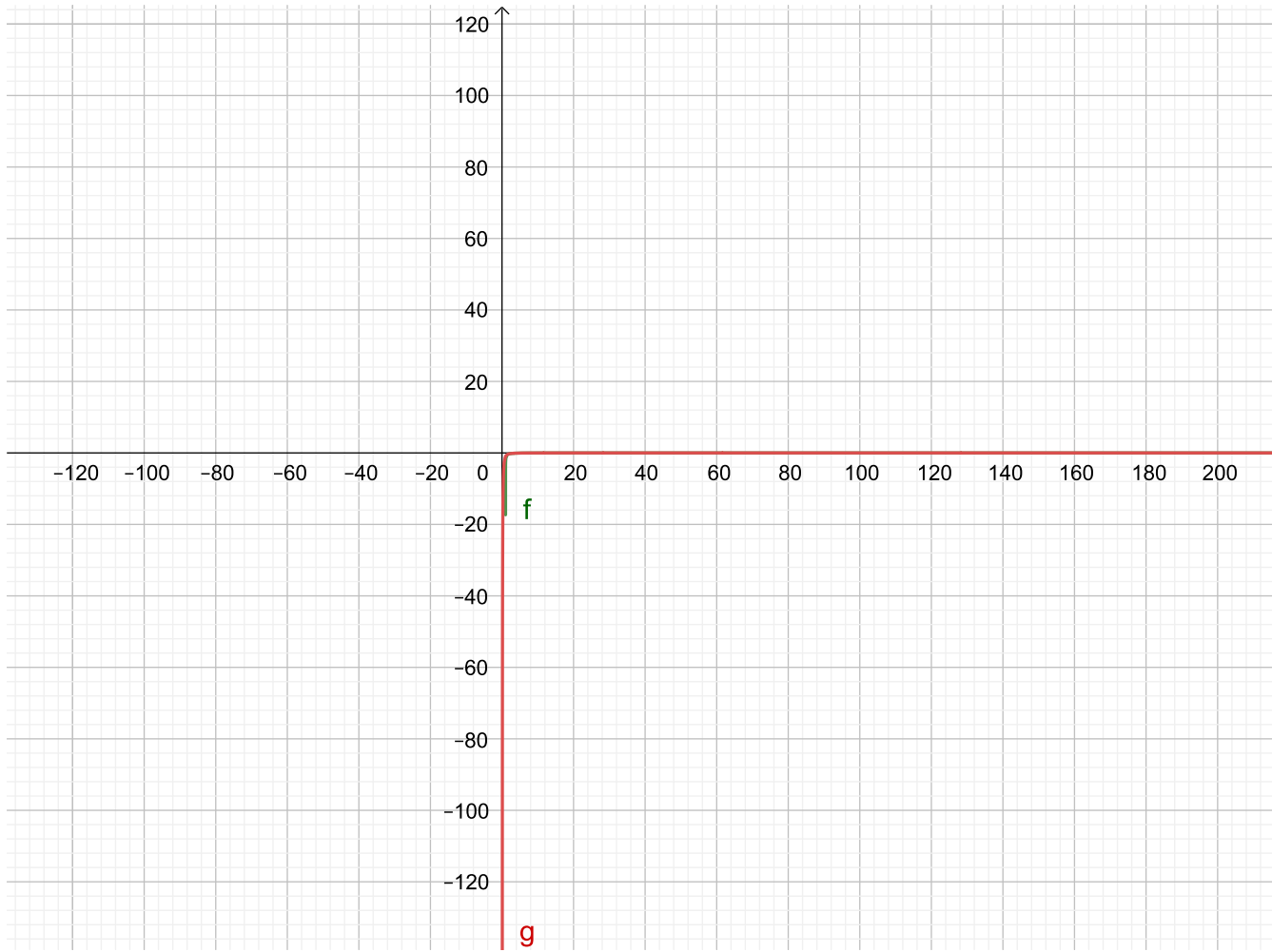}
\vspace{-0.2cm} \caption{Here we plot $F_e/C$, $C=\frac {3Nk_B\lambda} {8\pi}$.
Black curve $f$ (barely visible, very near the negative vertical axis): Fermion  entropic force. Note the cut-off near the origin.
Red curve $g$: approximate semi-classic Fermion-one.
Both curves  are almost-coincident for $r-$ not too small. For  better grasping details near the origin,  see Fig.2.
}\label{fig1}
\end{center}
\end{figure}

\begin{figure}[h]
\begin{center}
\includegraphics[scale=0.5,angle=0]{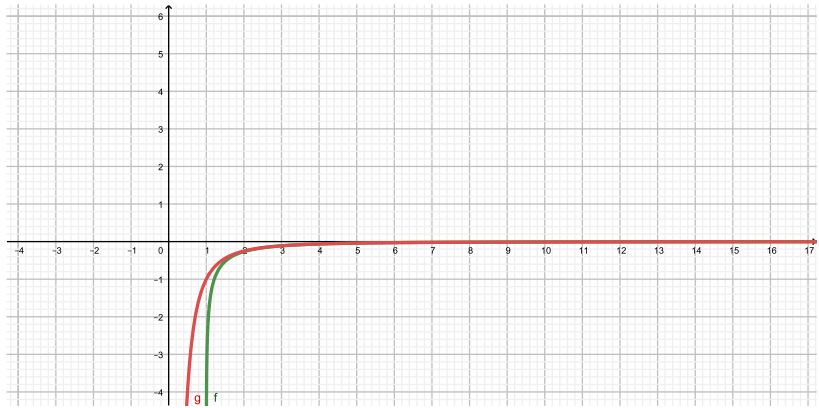}
\vspace{-0.2cm} \caption{Here we plot an  amplified part, near the origin,  of Fig. 1.}\label{fig2}
\end{center}
\end{figure}

\nd In Fig. 1 we plot  $F_e$ with $m$= uranium's atom mass.
$E=\frac {Nmv^2} {2}$, $v=1000\,\, meter/second$, and  $N=500$
yielding $r^3>1.525\times 10^{-33}$

\nd The cut-off originates because $F_e$ takes now  the simplified form
\begin{equation}
\label{eq2.15}
F_e=\frac {3N\lambda k_B} {8\pi}r
\ln\left[1-\frac  {1.525\times 10^{-33}} { r^3}\right]
\end{equation}
and the plot is drawn for
\begin{equation}
\label{eq2.16}
\frac {F_e} {\left(\frac {3N\lambda k_B} {8\pi}\right)(1.525\times 10^{-33})^{\frac {1} {3}}}=x
\ln\left[1-\frac  {1} { x^3}\right].
\end{equation}


\section{Conclusions}

\nd We have considered,  for fermions,   Verlinde's [entropic force - Gravitation] link, proved recently both in a classical context  \cite{p1} and in a quantum bosonic scenario \cite{DOI}.  We have seen that Verlinde's conjecture holds also for  a Fermi environment. Further, the quantum emergent gravitation \`a la Verlinde {\it does not diverge at the origin} because a cut-off impedes reaching it. 

\nd One finds, however, the emergent  gravitation's usual divergence-at-the-origin in the classical limit.  One might perhaps wonder whether this divergence could be a  classical artifact, since 
for bosons one does not have divergence at the origin neither \cite{DOI}.  Note that  in two limits

\begin{itemize}

\item the classical limit of QM  
\item   $r $ large enough.

\end{itemize}
the Newton $r$-dependence of the gravitation force is recovered. 
\vskip 3mm

\nd Finally, we  remark that we find an  $r-$cut-off in the fermionic entropic force that somehow 
becomes  reminiscent of the space-discretization ideas of loop gravity.  
\cite{loop}. Our approach is not yet able to deal with geometric facets \`a la Einstein but might be coherent with granular 
space-time.

\setcounter{equation}{0}

\vskip 5mm 

\section*{Acknowledgements}

\nd We are most grateful to Prof. A. R. Plastino for useful discussions.


\newpage




\begin{thebibliography}{99}


\bibitem{verlinde} E. Verlinde,  arXiv:1001.0785 [hep-th];
JHEP {\bf 04}, 29 (2011).

\bibitem{p1} A. Plastino, M. C. Rocca: Physica A {\bf 505}, 190 (2018).

\bibitem{DOI}  \url{https://www.researchgate.net/publication/324784383_Quantum_treatment_of_Verlinde's_entropic_force_conjecture}
DOI10.13140/RG.2.2.34454.24640.


\bibitem{times} D. Overbye,   {\it A Scientist Takes On Gravity}, The New York Times, 12 July 2010;
M. Calmthout, New Scientist {\bf 205},6   (2010).

\bibitem{libro} J. Makela, arXiv:1001.3808v3;
J. Lee, arXiv:1005.1347;
V. V. Kiselev, S. A. Timofeev ,
Mod. Phys. Lett. A {\bf 25}, 2223 (2010);
T. Aaltonen et al;
Mod. Phys. Lett. A {\bf 25}, 2825  (2010).

\bibitem{prd}  A. Sheykhi, Phys. Rev. D {\bf 81}, 104011 (2010); 
S. Hossenfelder, Phys. Rev. D {\bf 95}, 124018 (2017) ;   
De-Chang Dai,  D. Stojkovic, Phys. Rev. D {\bf 96}, 108501 (2017).


\bibitem{india} T. Padmanabhan, arXiv 0812.2610v2.

\bibitem{guseo} R. Guseo, Physica A {\bf 464}, 1 (2016).


\bibitem{lemons} D. S. Lemons,{\it A Student's Guide to Entropy},
Cambridge University Press (Cambridge, UK, 2014).

\bibitem{loop} Scientific American Time,
{\bf 103}, 94  (2012); C. Rovelli, {\it Quantum Gravity}, 
\url{http://arxiv.org/abs/gr-qc/9710008v1}, and references therein

\bibitem{gra} I. S. Gradshteyn and I. M. Ryzhik : ``Table of Integrals,
Series and Products''. Academic Press, Inc (1980).


\end{thebibliography}
\end{document}